\documentstyle[amssymb,aps,prl]{revtex}

\begin{document}
\author{D.Mihailovic$^{\dagger }$, T.Mertelj$^{\ddagger \dagger }$}
\address{$^{\dagger }$Solid State Physics Department, Jozef Stefan Institute, 1001\\
Ljubljana, Slovenia}
\address{$^{\ddagger }$Faculty of Mathematics and Physics, Jadranska 19, 1000\\
Ljubljana, Slovenia}
\author{K.A.M\"{u}ller}
\address{Physics Department, University of Zurich, CH-Zurich, Switzerland}
\title{$a-b$ plane optical conductivity in YBa$_{2}$Cu$_{3}$O$_{7-\delta }$\ above
and below $T^{*}$.}
\maketitle

\begin{abstract}
Analysis of the $a-b$ plane optical conductivity $\sigma _{ab}$ for both
twinned and untwinned YBa$_{2}$Cu$_{3}$O$_{7-\delta }$ as a function of
temperature and doping shows that below a well defined temperature $T^{*},$
a dip in the spectrum systematically appears separating the infrared charge
excitation spectrum into two components with distinct energy scales. The
change from monotonic behaviour in $\sigma _{ab}$ is found to be concurrent
with the onset of phonon anomalies in Raman and infrared spectra below $%
T^{*} $. The optical data are suggested to be evidence for the appearance of
an inhomogeneous distribution of carriers rather than the opening of a
simple gap for charge excitations below $T^{*}$, an interpretation wich is
consistent with recent angle-resoved photoemission and electronic Raman
spectra. We find that the behaviour below $T^{*}$ and the absence of any
anomalies at $T_{c}$ can be interpreted assuming a Bose-Einstein
condensation of preformed pairs.
\end{abstract}

Following theoretical ideas on inhomogeneous charge carrier distributions
and phase separation in cuprate superconductors\cite
{EmeryKivelson,SigmundHiznaykov,DiCastro}, significant experimental support
has recently been accumulating for a two-component charge carrier picture,
particularly in the underdoped region of the phase diagram of these
materials. Structural evidence from extendend X-ray absorption fine
structure (EXAFS)\cite{Bianconi}\ and neutron pair distribution function
analysis (PDF)\cite{Egami}\ as well as NMR, NQR\ and Mossbauer\cite
{NMR,Teplov,NQR,Mossbauer} spectroscopies\cite{MihailovicMueller} indicate
that below a certain cross-over temperature $T^{*}$, the Cu-O planes may
contain stripes of carrier-rich and carrier-poor regions on length scales
typical of the superconducting coherence length $\xi _{s}$. These features
apparently have quite fast dynamics\cite{Stevens} making the textures
observable only with spectroscopies with sufficiently short characteristic
timescales. Although initially the NMR data suggested that the underdoped
state may be characterized by a gap primarily for spin excitations, recent
angle-resolved photoemission (ARPES)\cite{ARPES} and time-resolved optical
photomodulation data\cite{Rome} suggest that some kind of a gap may also be
present - at least for parts of the Fermi surface - for charge excitations
as well. The appearance of such a pseudogap above $T_{c}$ appears to be
quite a general phenomenon and is observed also in many other systems,
including the superconducting bismouthates like Ba$_{1-x}$K$_{x}$BiO$_{3}$%
\cite{BKBO}, heavy Fermion superconductors\cite{Steglich} and possibly even
in the roton excitation spectrum of LHe\cite{LHe}, implying a possible
common underlying reason for its existence in all these systems.

However in the case of the cuprates, the data from a key experiment -
infrared optical spectroscopy - has not yet been systematically discussed in
terms of the above-mentioned picture. Although the existence of a
''pseudogap'' in the out-of-plane optical conductivity $\sigma _{1}^{c}$ has
been investigated in some detail\cite
{HomesTimusk93,BasovTimusk94,HomesTimusk95pseudo,Tajima96}, the in-plane
optical data has so far been discussed mainly in terms of a single-component
framework of an extended Drude model with temperature- and
frequency-dependent scattering rate $\tau (\omega ,T)$ and carrier mass $%
m^{*}(\omega ,T)$\cite{PuchkovFournier96,PuchkovBasov96} and no systematic
analysis of $\sigma (\omega ,T)$ as a function of doping {\em and}
temperature has yet been performed.

It is the purpose of this paper to present an analysis of the available $a-b$
plane optical conductivity data in terms of the two-component picture. We
concentrate primarily on YBa$_{2}$Cu$_{3}$O$_{7-\delta }$ where the data are
most complete and the contributions to the optical conductivity arising from
the chains and planes can be identified from $a-b$ anisotropy measurements
of $\sigma _{1}(\omega ,T)$\cite{RotterSchlessinger91}. Our analysis of the
doping and temperature dependence of the optical conductivity are performed
on published data and unambiguously show that the in-plane optical
conductivity qualitatively changes at a characteristic temperature $T^{*}$
from a single-component spectrum above, to one with (at least) two
components below $T^{*}$. These findings together with enumeration of
various phonon anomalies in the Raman and infrared spectra, collected from
the available data apparently confirm the two-component paradigm of the
normal state in the underdoped materials below $T^{*}$, and define the
temperature and frequency dependence of the excitations in the two-component
state.

In common with most near-optimally doped cuprate samples, the real part of
the room temperature optical conductivity along the crystallographic $a$%
-axis $\sigma _{1a}$ of YBa$_{2}$Cu$_{3}$O$_{7-\delta }$ below $\omega \sim
400$ cm$^{-1}$ shows a monotonic decrease with increasing frequency
reminiscent of an inverse law $\sigma (\omega )\sim 1/\omega $ $.$ Above
this frequency the inverse law fails and an additional spectral weight is
necessary to describe the spectra\cite{Timusk}. Below room temperature (but
still above $T_{c}$) a dip ubiquitously appears in the range 400 - 800 cm$%
^{-1}$ which becomes more pronounced with further cooling and eventually
below $T_{c}$ the residual conductivity shows only a characteristic
mid-infrared (MIR) peak at around 1000 cm$^{-1}$ which again is a feature
common to all the cuprates. To analyse this behaviour of $\sigma _{1}(\omega
,T)$ more quantitatively, we have selected an arbitrary criterion to define
the temperature $T^{*}$ at which $\sigma _{1}(\omega ,T)$ departs from
monotonic behaviour and the first derivative $\sigma ^{\prime }(\omega )$
shows a maximum (i.e. a point of inflexion appears) as shown schematically
by the arrow in the Figure 1. Data for a number of YBa$_{2}$Cu$_{3}$O$%
_{7-\delta }$ samples with different $T_{c}$'s and $\delta $ were collected
and the results of the analysis are plotted as a function of $T_{c}$ in
Figure 2. Because there are usually not many spectra at different
temperatures available, the data are necessarily plotted with a range of
temperatures indicated as error-bars. Nevertheless, the correlation between
the temperature $T^{*}$ where the spectrum departs from monotonic behaviour
and $T_{c}$ is quite evident and the relation appears almost linear in YBa$%
_{2}$Cu$_{3}$O$_{7-\delta }$ over the whole range of $\delta $.

To complement the analysis on $\sigma _{1}$, we have systematically
collected infrared and Raman phonon spectra on YBa$_{2}$Cu$_{3}$O$_{7-\delta
}$ with different $\delta $ and determined the temperatures $T_{anomaly}$ at
which phonon anomalies appear. To expand the data set \thinspace somewhat we
have also used the published data on phonon anomalies in the 124 and
\thinspace 247 structures, which have similar optimal $T_{c}$ as 123\cite
{KaldisKarpinsky91}. IR phonons anomalies are observed above $T_{c}$ for
phonons involving $c$-axis displacements of Y, Ba, O1 (chain), O2/O3, O4 and
Cu1(chain) ions\cite{ShutzmanTajima95,HomesTimusk95} in 123 system and of O1
(chain) and O2/O3 ions in 124 system.\cite
{LitvinchukThomsen92,LitvinchukThomsen93}. Among Raman active phonons it is
interesting to note that the 340-cm$^{-1}B_{1g}$-like O2/O3 planar buckling
phonon shows little or no anomalies above $T_{c\,}$ in underdoped region of
doping.\cite{LitvinchukThomsen92,KrantzRosen88,HeyenCardona91} On the other
hand the related in phase $A_{g}$ O2/O3 planar phonon at 440 cm$^{-1}$ shows
a significantly more pronounced anomaly above $T_{c}$.\cite
{LitvinchukThomsen92,HeyenCardona91} In the 124 system\cite{HeyenCardona91}
anomalies above $T_{c}$ in underdoped region are reported also for the 108-cm%
$^{-1}$Ba phonon, 256-cm$^{-1}$ Cu1 phonon and 501-cm$^{-1}$ O4 phonon.
There is no clear data on whether such anomalies for Ba and O4 phonons exist
also in YBCO-123.

The citerion for determination of the temperatures $T_{anomaly}$ was a
discontinuous change in the phonon frequency and/or linewidth. A plot of $%
T_{anomaly}$ as a function of doping is shown in the insert to Figure 2 as a
function of $T_{c}$. The larger scatter in the data in this case is partly
due to the ambiguity in determining the exact temperature where a phonon
anomaly occurs. Despite this, it can be seen, that in virtually all cases
phonon anomalies appear below $T^{*}$. Depending on $\delta ,$ $T_{anomaly}$
seems to vary such that in the ortho-II phase of YBCO (70K $<T_{c}<$\ 93K), $%
T_{anomaly}$ $\lesssim T^{*}$ , while in the Ortho-I phase (0 $<T_{c}<$\ 70
K), $T_{anomaly}$ $\ll T^{*}.$ For optimal doping there is no sign of any
phonon anomalies above $T_{c},$ so we assume $T_{anomaly}\thickapprox T_{c}$%
. Unfortunately, at present there is insufficient data to be able to
correlate the behaviour of particular phonons with different values of $%
T_{anomaly}$ and there is a clear need for such a study in the future.

For completeness, the $T^{*}$ data from optical conductivity are plotted as
a function of oxygen content in Figure 3\cite{explanation} together with $%
T_{c}$. These data for $T^{*}$ obtained from $\sigma _{ab}$ are in good
agreement with data on $\sigma _{c}$\cite
{HomesTimusk93,BasovTimusk94,HomesTimusk95pseudo,Tajima96} and with other
data from a variety of experimental techniques on YBCO, including ARPES, DC
transport and NMR, where it is often called a spin gap temperature $T_{sg}$.
In the present case, the optical conductivity experiments measure only
dipole transitions, and $T^{*}$ therefore signifies the pseudogap opening
for charge excitations. The appearance of a pseudogap for spin excitations
will thus necessarily follow at the same temperature. (The converse is is
not necessarily true). Although there appears to be an apparent spread of
values of $T^{*}$ and $T_{sg}$ in the literaure, suggesting the possible
existence of {\em two }pseudogaps appearing at different temperatures, when
it is taken into account that very different criteria are used to define $%
T^{*}$ by different authors, and that different spectroscopic techniques
average over momentum in different ways, data on YBCO do not appear to
support the two gap hypothesis at present. Careful examination of the
optical conductivity data for $\sigma _{ab}$ and $\sigma _{c},$ particularly
in comparison with the data on the increase in NMR\ $T_{1}$ relaxation time
(usually expressed as a drop in $1/T_{1}T$)\cite
{T1relaxation,TakigawaReyes91}, NMR Knight shift $K$\cite
{TakigawaReyes91,Knightshift}$,$ neutron scattering\cite{neutron} and
specific heat data\cite{Loram} does not yield any systematic difference in $%
T^{*}$. However, with significantly improved sample characterization in the
underdoped state and a well-defined common criterion for defining $T^{*}$
the existence of two distinct, closely spaced gaps cannot be excluded.

Let us now turn to a discussion of the presented data in the general
framework of the inhomogeneous phase models. Commonly, the state below $%
T^{*} $ is suggested to consist of carrier-poor and carrier-rich regions.
The optical response is expected to reflect this, the existence of
identifyably different spectral features below $T^{*}$ and a single
component above $T^{*} $ is in clear accordance with these models. Below $%
T^{*}$, the low-frequency spectrum - which is assumed to be due to itinerant
charges - is strongly temperature-dependent, has no identifyable spectral
signatures and is described well phenomenologically by an extended Drude
model (albeit over a small spectral region). As a corrollary, the MIR would
be expected to be due to the excitation of bound localized carriers {\em to
the itinerant carrier 'Fermionic' phase} in addition to non-adiabatic
hopping processes within the insulating phase itself as will be discussed
below.

Examining the behaviour of the MIR feature as a function of $\delta $ in
more detail, we have plotted the residual conductivity at the peak in the
MIR (usually in the range 1000 - 1300 cm$^{-1}$) below $T_{c}$ as a function
of doping in Figure 4. The data extracted from Rotter {\it et al. }\cite
{RotterSchlessinger91} are of particular relevance because they are from 
{\em untwinned} single crystals (also for lower $\delta $). We find an
unambiguous increase in $\sigma _{1}^{MIR}$ with increasing $\delta $.
Comparing the $\sigma _{1a}^{MIR}$ with $\sigma _{1ab}^{MIR}$ shows that the 
{\em in-plane }conductivity $\sigma _{1ab}^{MIR}$ shows very similar
systematics as the total conductivity $\sigma _{1ab}^{MIR}$. The ratio $%
\sigma _{1ab}^{MIR}/\sigma _{1a}^{MIR}$ $=\zeta $ appears to be only weakly
dependent on $\delta $, indicating that the increasing O doping gives rise
to increasing conductivity in the chains and planes on an almost equal
footing. Moreover, the ratio between the total and chain conductivity is $%
\zeta =1.4$ to $1.8$ and is surprisingly similar to the anisotropy of the DC
conductivity $\sigma (0)$\cite{DCconductivity} which suggests that O doping
results in an increase in conductivity of {\em both} the Drude and MIR\
parts equally (see Figure 4).

The spectral shape of the MIR is similar (but not identical) to the
photoinduced polaronic features observed in the insulating phase in YBCO,
and is virtually indistinguishable from the MIR\ response of the insulator
PrBa$_{2}$Cu$_{3}$O$_{7-\delta }$\cite{Takenaga,KircherHumlichek91}
(including anisotropy). This, together with clearly visible phonon overtones%
\cite{calvani} in the MIR\ spectrum led many authors in the past to a
(bi)polaronic interpretation of the MIR\cite{calvani,MihailovicFoster90}. In
this model, the MIR absorption results from non-adiabatic carrier hopping
processes where a different number of phonons is emitted and reabsorbed, the
difference in energy being taken up by the IR photon. However, the measured
MIR\ spectral shape shows significant deviations from the predicted
polaronic spectrum\cite{MihailovicFoster90,Reik} and expansion of this
simple polaron hopping model to include a realistic density of phonon states
and allowing for phonon dispersion still cannot reproduce the highly
asymmetric peak in the MIR data. However, when we consider a two-component
state, an additional process becomes possible, namely the excitation of
carriers from the polaronic bound state to the unbound Fermionic state\cite
{Kabanov}. The expected MIR\ spectrum in this case exhibits an asymmetric
spectral shape much closer to the experimentally observed one$.$ A more
sophisticated attempt at describing the MIR\ lineshape in detail might
include both excitations from the localized states to the delocalized band%
\cite{Kabanov} as well as non-adiabatic\cite{Reik} hopping processes between
localized states.

The interpretation of the MIR given above is supported by recent resonant
electronic Raman spectra of Ruani and Ricci\cite{Ruani}, where an electronic
peak in the region of 0.06 eV (500 cm$^{-1}$) is observed below $T^{*}.$ The
authors attribute this peak to excitations between dispersionless polaronic
bands near the S-point of the Brillouin zone. Further support for the
proposed interpretation comes from the observation of a depression in the
density of states below $T^{*}$ in the same region of the Brillouin zone in
ARPES\cite{ARPES}: to satisfy the sum rule, (bi)polaron formation
(i.e.pairing) below $T^{*}$ must be accompanied by a depletion in the
density of states near $E_{F}$, causing the appearance of a pseudo-gap in
the ARPES\ spectrum.

The present analysis is essentially an alternative phenomenological view of
the electrodynamic response of the cuprate superconductors to the
one-component memory function parametrization more usually discussed in the
recent past\cite{PuchkovFournier96,PuchkovBasov96}. We believe it provides
additional physical insight into the normal and superconducting state
properties in line with the numerous recent suggestions for the co-existence
of two {\em structural }local units in these materials\cite
{MihailovicMueller}. Analysis of the optical conductivity in YBa$_{2}$Cu$%
_{3} $O$_{7-\delta }$ as a function of $\delta $ and $T$ reveals some
seemingly universal features which could be considered as a possible
definition of the low-energy charge excitation spectrum for the cuprates:

\begin{enumerate}
\item  The $a-b$ optical conductivity $\sigma _{1ab}$ or $\sigma _{1a}$
(with or without chains respectively) in YBCO universally displays a break
from monotonic behaviour near 400 cm$^{-1}$ (50 meV) below some temperature $%
T^{*}.$ The changes in the optical spectrum below $T^{*}$ suggest a
separation of $\sigma _{1}$ into two separate responses with characteristic
excitation energy scales and well-defined and different $\omega $ and $T$%
-dependences, rather than the opening of a simple gap. Above $T^{*}$ the two
excitations appear indistinguishable, presumably as a result of thermal
fluctuations.

\item  Below $T^{*}$ the low-frequency optical conductivity ($\omega <400$ cm%
$^{-1}$) of the itinerant carriers is apparently described reasonably well
by an extended Drude model with a frequency- and temperature-dependent
scattering rate $\tau _{s}$ \cite{PuchkovBasov96}. Assuming a Fermi energy
of 0.1 eV and $m^{*}=4m_{e}$, we find that the mean free path $\lambda _{c}$ 
$=v_{F}\tau _{s}$ for the Drude carriers at $T^{*}$ is $\sim 25$ \.{A}\
which is remarkably close to the typical length scale of the spatial
inhomogeneities (stripe widths) found in structural experiments\cite
{Bianconi,Egami}, suggesting that the properties of the Drude carriers are
determined by the physics of the Fermions in the confined structure. The
increasing $\lambda _{c}$ at lower temperatures might be a sign of
increasing coherence between stripes, rather than increasing stripe width.

\item  The 2-component analysis is consistent both with the observations of
a ''pseudogap'' in transport\cite{Batlogg}, NMR\cite{NMR Tallon} and ARPES%
\cite{ARPES}. An explanation consistent with all four experimental
techniques involves the formation of polarons in carrier-poor regions below $%
T^{*}$ giving rise to a reduction in the DOS at $E_{F}$ and hence a drop in
the $\chi (\omega =0)$ for NMR, as well as an apparent shift in the binding
energy $E_{b}$ in ARPES. This polaronic effect is indeed expected to be
associated with the flat, dispersionless regions of the Fermi surface,
consistent with the enhanced polaron effective mass $m^{*}$, the electronic
Raman\ data\cite{Ruani} and ARPES\cite{ARPES}.

\item  The in-plane MIR conductivity $\sigma _{1a}^{MIR}$ - which is
presumed to be principally due to the excitation of localized carriers to
the Fermionic regions in the Cu-O planes - increases in intensity
systematically with $\delta $ closely following the density of doped holes
and is only weakly temperature dependent.

\item  Numerous phonon anomalies appear in the optical conductivity in a
range of temperatures such that $T_{anomaly}$ $<T^{*}$, suggesting a
coupling of the lattice degrees of freedom to the electronic charge
excitations. Unfortunately there is insufficient data available at present
for a more detailed examination of which phonons are involved, but the
structural data from EXAFS\cite{Bianconi}\ and PDF\cite{Egami} appear to
corroborate this observation. The $T^{*}$ observable for charge excitations
appears to be consistent with $T_{sg}$ in NMR and other experimental
techniques, and is at this time experimentally indistinguishable from it.
\end{enumerate}

Finally, we briefly discuss the possible involvement of (bi)polarons in
superconductivity. Their role was usually dismissed as unimportant on the
grounds that there is no observable change in $\sigma (\omega )$ at MIR
frequencies at $T_{c}$. In the BCS\ scenario, pairing occurs simultaneously
with the opening of a gap in the charge excitation spectrum and the
appearance of a macroscopic phase coherent state at (or very near) $T_{c}$.
As a result of the gap, a significant change in spectral weight is observed
in $\sigma (\omega )$ below $T_{c}$. In contrast, in the BE condensation
picture, pairing is an independent process from condensation and the
establishment of phase coherence. Of course it must occur above (or at) the
condensation temperature because the process of condensation requires a
macroscopic number of particles with Bose statistics to be present. In the
case of underdoped cuprates, where neither the Drude nor MIR show any change
at $T_{c}$ itself, the implication is that BE condensation of pre-formed
pairs takes place with no change in pairing amplitude. It is natural to
presume that because the redistribution of spectral weight in the optical
spectrum starts to occur at $T^{*},$ this temperature should be associated
with the onset of pair formation. Hence we conclude the paper by pointing
out that the suggested two component interpretation of the optical
conductivity spectra in underdoped cuprates is consistent with pair
formation at $T^{*}$ and the Bose-Einstein condensation of such pairs
across, or together with itinerant 'Drude' carriers into a macroscopically
coherent ground state at $T_{c}$ \cite{Ranninger}.

\section{Acknowledgments}

The work was partly supported by the EU. One of us (D.M.) wishes to
acknowledge discussion with D.Van derMarel and V.V.Kabanov.

\section{Figure Captions}

Figure 1. A schematic of the optical conductivity $\sigma _{1a}$
perpendicular to the chains in YBCO at different temperatures. Similar
behaviour is observed also in Bi and La cuprate compounds.\cite
{OrensteinThomas90}

Figure 2. $T^{*}$ as a function of $T_{c}$ as determined from the appearance
of an inflexion point in the in-plane optical conductivity. Seemingly there
is no significant difference between $\sigma _{_{1}ab}$ and $\sigma _{1a}$.
Insert: $T_{anomaly}$ as a function of $T_{c}$ as determined from phonon
anomalies in the infrared and Raman spectra of 123, 124 and 247. Full
symbols represent Raman data.

Figure 3. The phase diagram of YBCO determined from the in-plane optical
conductivity. Seemingly there is no significant difference between $\sigma
_{_{1}ab}$ and $\sigma _{1a}$.

Figure 4. The magnitude of the ''residual'' MIR as a function of a) O
content and b) $T_{c}$ in YBaCuO. The upper data are for $\sigma _{1ab},$
while the lower data set are for $\sigma _{1a}$.


\begin{references}
\bibitem{explanation}  Some ambiguity arises in assigning $\delta $ from the
published $T_{c}$'s near optimum doping because of the existence of two
phases above $\delta =6.92.$ The exact point of intersection between the $%
T^{*}$and $T_{c}$ curves is ambiguous $\delta $ near optimum doping as
indicated by the two dotted lines.
\end{references}
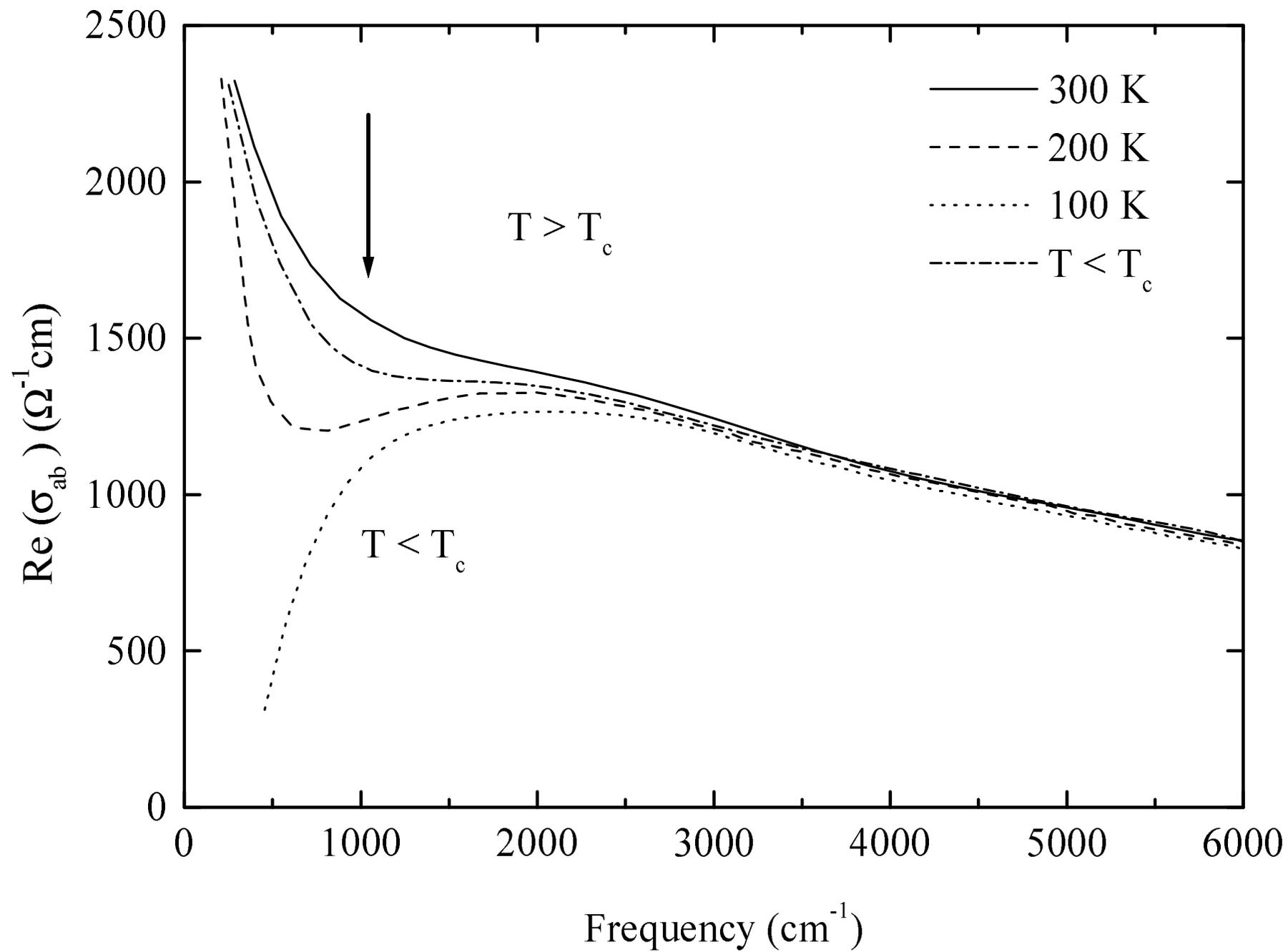
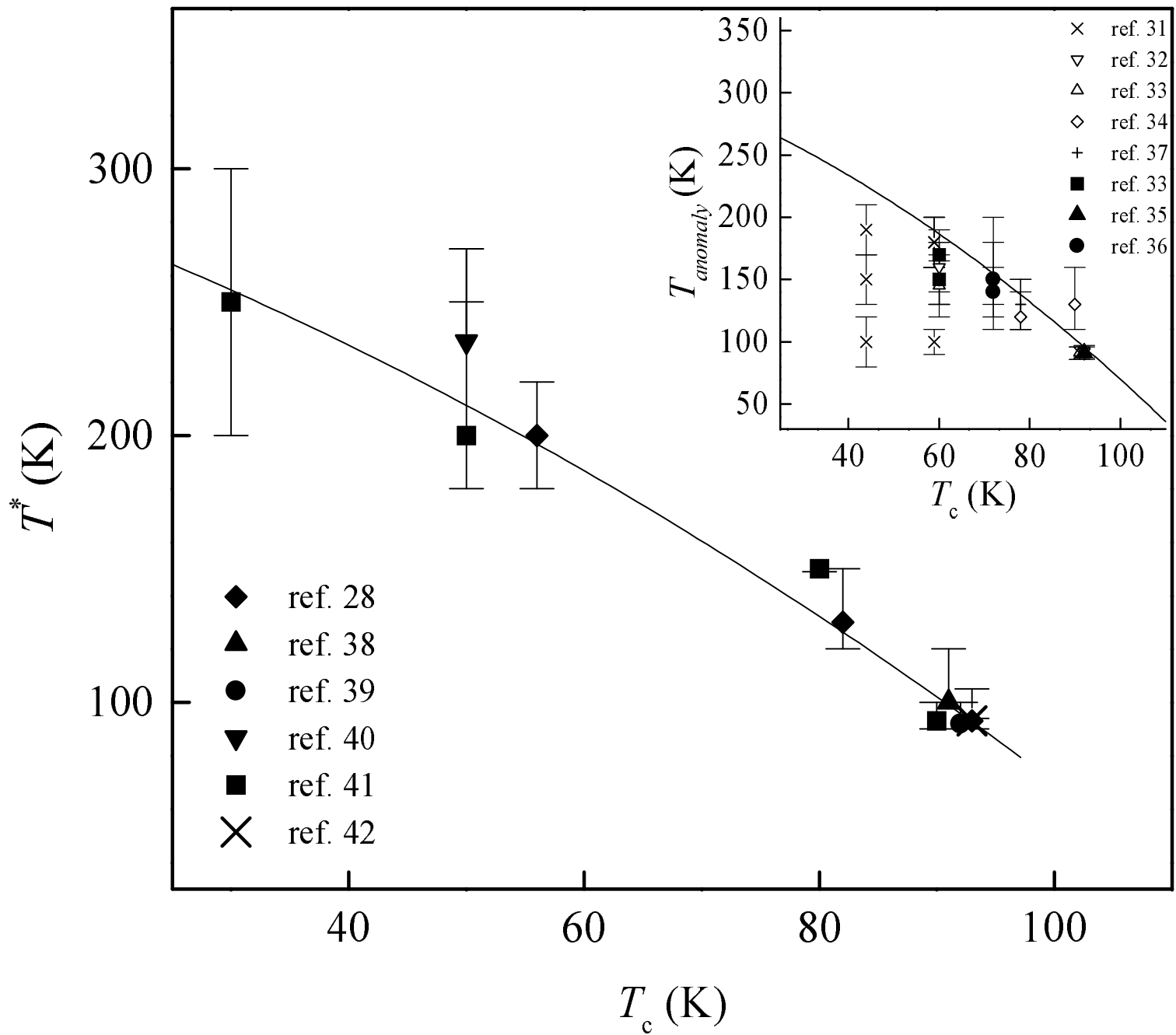
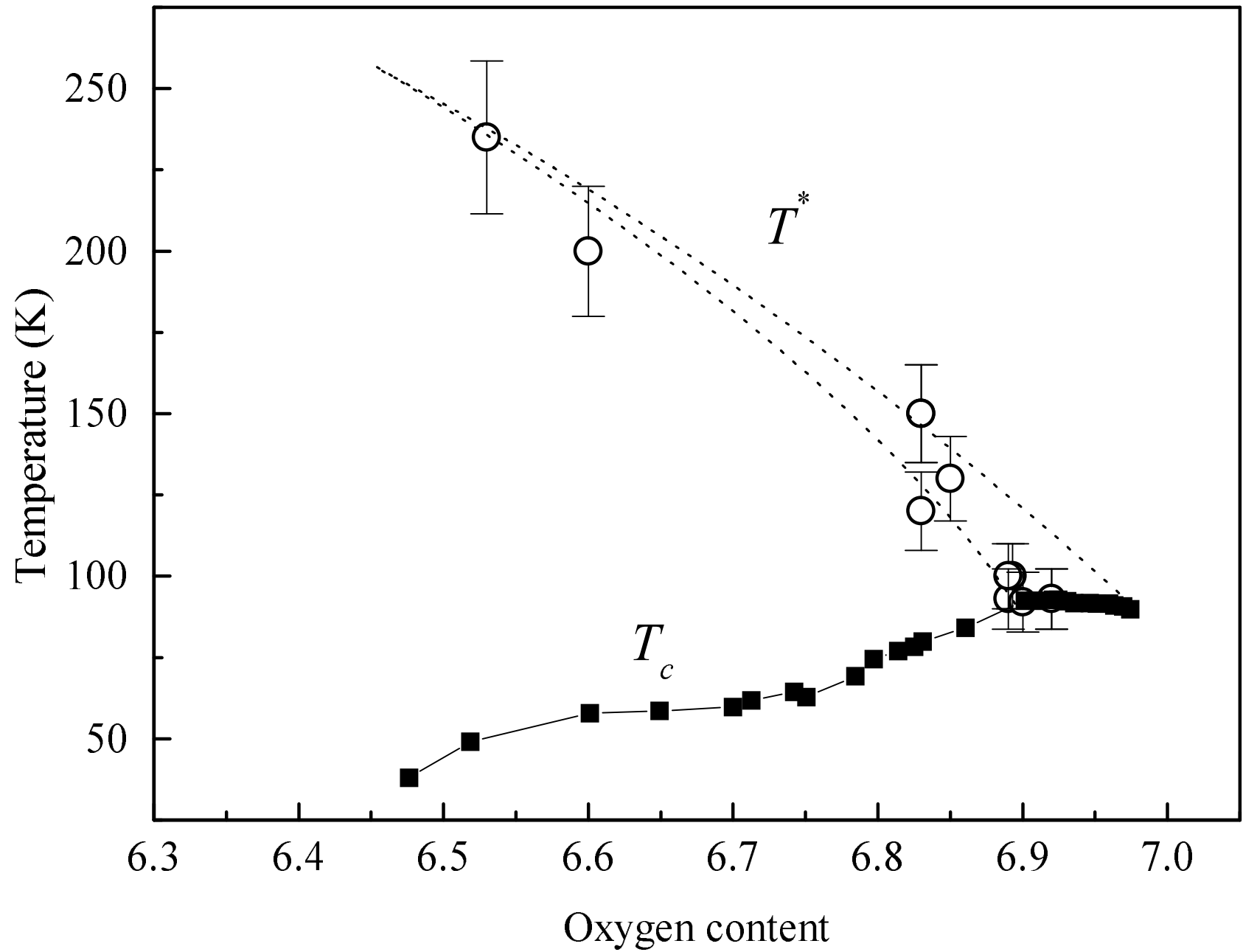
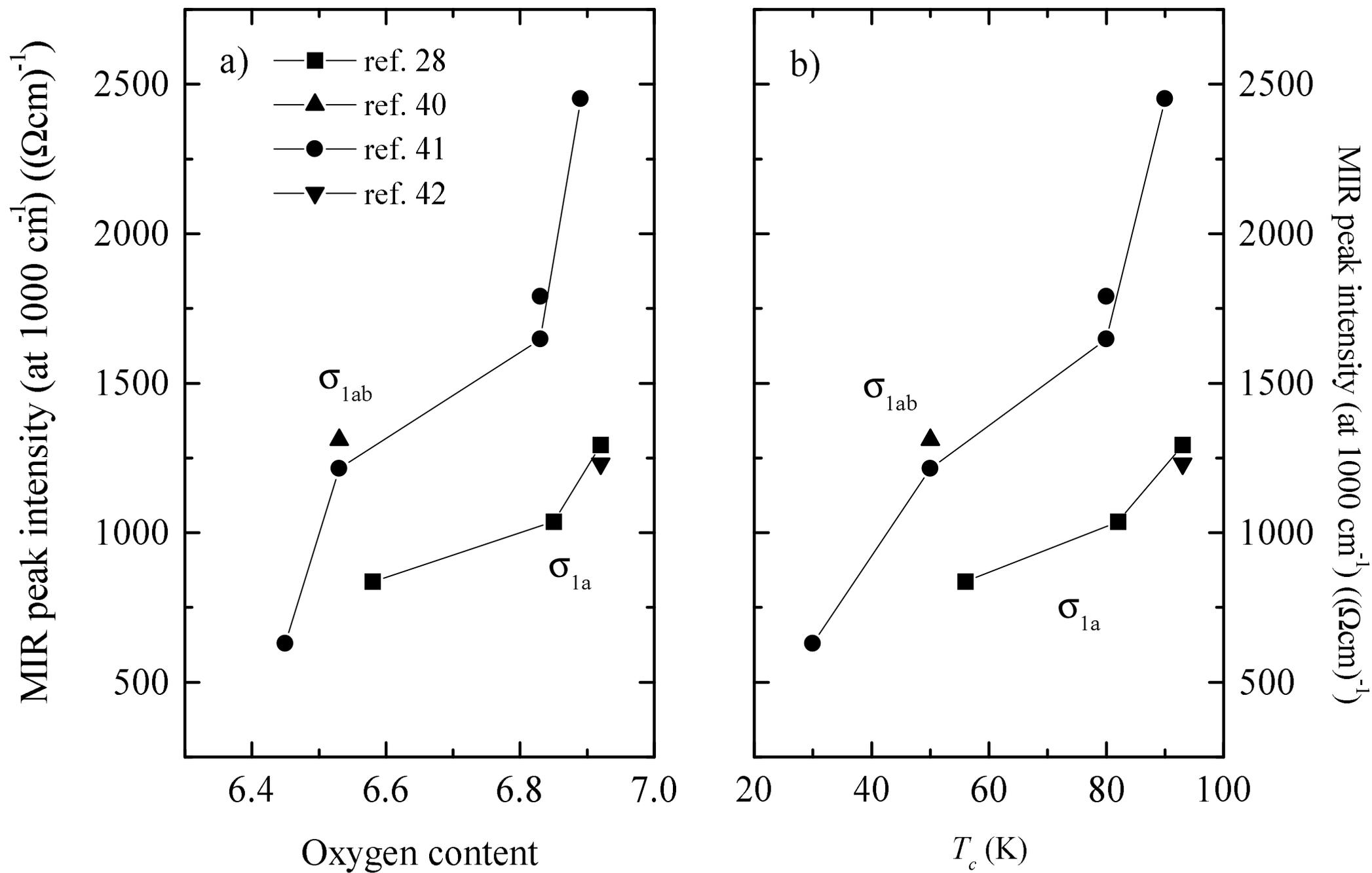
\end{document}